# Combinatorial discovery of a lead-free morphotropic phase boundary in a thin-film piezoelectric perovskite

S. Fujino[1], M. Murakami[1]*, A. Varatharajan[2]*, S.-H. Lim[1]*, V. Nagarajan[2], C. J. Fennie[3], M. Wuttig[1], L. Salamanca-Riba[1], and I. Takeuchi[1]

*These authors contributed equally to the work

[1]*Department of Materials Science and Engineering, University of Maryland, College Park, Maryland 20742 USA*

[2]*School of Materials Science, University of New South Wales, Sydney NSW 2052, Australia*

[3]*Materials Science Division, Argonne National Laboratory, Argonne, Illinois 60439, USA*

## Abstract

**We report on the discovery of a lead-free morphotropic phase boundary in Sm doped $BiFeO_3$ with a simple perovskite structure using the combinatorial thin film strategy. The boundary is a rhombohedral to pseudo-orthorhombic structural transition which exhibits a ferroelectric (FE) to antiferroelectric (AFE) transition at approximately $Bi_{0.86}Sm_{0.14}FeO_3$ with dielectric constant and out-of-plane piezoelectric coefficient comparable to those of epitaxial (001) oriented $Pb(Zr,Ti)O_3$ (PZT) thin films at the MPB. The discovered composition may be a strong candidate of a Pb-free piezoelectric replacement of PZT.**

BiFeO$_3$ (BFO) is a multiferroic with rich and intricate physical properties.[1] Given the need for environmentally friendly lead-free piezoelectrics[2], doped-BFO is being investigated for improved properties[3-5], so that they may emulate the performance of Pb-based piezoelectrics[6,7] at morphotropic phase boundaries (MPBs). However neither an increase in electromechanical constants as a function of dopant concentration or domain structures indicative of a MPB has been reported to date in BFO-based lead-free systems. There are some guidelines which predict the presence of MPBs[8-10] and one can explore novel compositions by systematically searching for similar structural transitions. Yet, comprehensive mapping of compositions requires synthesis of an enormously large number of individual samples. To overcome this challenge, we implement the combinatorial strategy in the form of composition spreads. This allows us to track precise changes in crystal structures as well as physical properties as a continuous function of composition in epitaxial thin film forms.[11]

A series of pseudo-binary composition spread epitaxial thin film libraries (200 nm thick) were fabricated on substrates 6 mm long by the combinatorial pulsed laser deposition system (Pascal, Inc.) at 600 °C, where a pseudo-binary compositional phase diagram of Bi$_{1-x}$(RE)$_x$FeO$_3$ or BiFe$_{1-y}$(TM)$_y$O$_3$ was continuously mapped on each chip. RE and TM denote rare-earth and transition metal cations, respectively. Composition variation across the spreads were confirmed by an electron probe (JEOL JXA-8900), and the uncertainty in the composition at each point on the spread is +/- 1.5%. Scanning x-ray microdiffraction (Bruker D8 Discover) was performed with a 0.5 mm diameter aperture. For electrical characterization, an epitaxial SrRuO$_3$ layer (50 nm) was used as the bottom electrode, and a sputter-deposited top Pd layer (50 nm) was patterned into 50 μm capacitor dots. Ferroelectric polarization hysteresis loops were obtained using the Radiant Precision LC at 5 kHz. Quantitative piezoresponse force microscopy was used to measure the out-of-plane piezoresponse. A number of interesting compounds were identified from our initial screening. Here, we focus on results from the $Bi_{1-x}Sm_xFeO_3$ (BSFO) spread.

X-ray diffraction (XRD) mapping of the (002) peak region in a BSFO composition spread on [001] $SrTiO_3$ (STO) are shown in Fig. 1(a). Initial piezoforce microscopy (PFM) scans of a BSFO spread indicated significant enhancement of piezoresponse near $Bi_{0.8-0.85}Sm_{0.2-0.15}FeO_3$, where the structure undergoes substantial change in the out-of-plane lattice constant. A detailed analysis of a series of two-dimensional XRD images of this composition region revealed appearance of extra diffraction spots for $x \geq 0.13$ (Fig.1(b)) indicating a cell doubling structural transition to a lower symmetry triclinic phase. Comprehensive structural analyses including detailed high resolution reciprocal space maps, electron diffraction patterns along various zone axes coupled with CARINE simulations will be published elsewhere. High temperature XRD of the same spread showed that the structural transition starts at approximately the same composition of $x \approx 0.14$, with no additional structural transitions up to the highest measured temperature of 400 °C.

High resolution planar transmission electron microscopy (A JEOL 2100 F operating at 200 kV) of an individual composition sample at $x \approx 0.14$ (Fig. 2) reveals presence of unusual nano-scale triclinic domains 20 – 50 nm in size displaying different relative epitaxial orientations. The lattice parameters of this composition were determined to be $a$ = 5.62 Å, $b$ = 7.83 Å, $c$ = 5.50 Å, $\alpha = \gamma$ = 89.8°, and $\beta$ = 89.7°. The occurrence of nanosized twins and concomitant stress accommodation have previously been identified as fingerprints of an adaptive ferroelectric phase at the MPB, which could result in a high piezoelectric coefficient and narrow hysteresis loops.[12] These nanodomains are observed only at this composition, and for x outside of 0.14 ± 0.015, the domains were found to be much larger in size (~100 nm).

Square-shaped ferroelectric hysteresis loops with good saturation and robust switchable polarization are obtained for compositions from x = 0 up to x ≈ 0.14. Figure 3 plots hysteresis loops at room temperature for three compositions (BFO, $Bi_{0.86}Sm_{0.14}FeO_3$ and $Bi_{0.84}Sm_{0.16}FeO_3$) selected for displaying the most prominent features. While increasing the Sm concentration induces a large drop in the coercive field, a high switchable polarization (≈ 70 $\mu C/cm^2$) is maintained. For x ≥ 0.15, we observe double hysteresis loops, indicative of an antiferroelectric (AFE) behavior.[13] It is important to note that the AFE composition at x ≈ 0.16 still maintains a relatively high polarization once electric-field induced switch to FE takes place. These hysteresis curves do not change their shapes with time or the number of measurement cycles, indicating that the double hysteresis loops are not from domain-wall pinning.

In Figure 4, we plot the zero-bias out-of-plane dielectric constant ($\varepsilon_{33}$) and loss tangent measured at 10 kHz as a function of increasing Sm concentration. The dielectric constant reaches a maximum at x = 0.14 in agreement with the structural transition taking place at the composition. The loss tangent at this composition is relatively low (~0.01). The inset shows the $\varepsilon_{33}$ versus electric field curve obtained for the $Bi_{0.85}Sm_{0.15}FeO_3$ composition displaying the characteristic multiple-peak behavior for an AFE material.

Figure 5(a) shows the piezoelectric behavior of BSFO with x = 0.12 and 0.14 measured via quantitative piezo force microscopy, which was performed on Pt/BSFO/SrRuO3 (SRO)/STO structures using Pt-Ir coated contact mode tips.[14] The measured $d_{33}$ values are effective values due to the constraint imposed by the underlying substrate. The composition right near the MPB possesses a substantially higher remanent out-of-plane $d_{33}$ (~95 pm/V) together with much reduced coercive field compared to that for x = 0.12. The remanent $d_{33}$ for a (001) oriented epitaxial PbZr$_{0.52}$Ti$_{0.48}$O$_3$ thin film with the same nominal thickness (200 nm) is 100 pm/V.[14]

Figure 5(b) plots the high-field $d_{33}$ as a function of Sm doping. Around x = 0.13 ~ 0.15, the effective $d_{33}$ displays a rapid increase peaking at x = 0.14 with 110 pm/V. Beyond this value, it rapidly decreases to ≈ 55 pm/V for Bi$_{0.83}$Sm$_{0.17}$FeO$_3$. The inset to the figure plots the Rayleigh analysis[15] of the measured out-of-plane piezoresponse for the MPB composition (Bi$_{0.86}$Sm$_{0.14}$FeO$_3$) sample. It shows that the magnitude of the piezoelectric coefficient scales linearly with the amplitude of the excitation sub-coercive ac electric field, while the second and third harmonic components are less than 1% of the total response even for high ac field. This behavior is quite different from the dynamic poling behavior observed in Pb-based piezoelectric thin films.[15] The lack of non-linear response in the first harmonic clearly suggests absence of any non-180º domain wall motion (pinned due to the substrate-induced constraint), and the near-zero second harmonic response indicates negligible electrostriction in this material.

The measured remanent and high field $d_{33}$ here are comparable to values previously reported for epitaxial thin films of Pb-based compounds such as PZT and PbMg$_{1/3}$Nb$_{2/3}$O$_3$-PbTiO$_3$. Domain engineered single crystals are known to exhibit enhanced electromechanical properties, but this is due to extrinsic mechanisms, which are generally absent in thin films [15]. Thus, in comparing nominally similar thin film samples of the same thickness, the MPB discovered here exhibits intrinsic piezoelectric properties which are among the best. The added advantage of the present system is a simpler crystal chemistry than some of the reported Pb-free compounds as well as ease of processing.[2]

There is a distinct and abrupt change in the $d_{33}$ loop shape as one crosses from the FE to AFE composition. Fig. 5(c) shows the *$d_{33}$* loop taken for $Bi_{0.84}Sm_{0.16}FeO_3$. We can relate the dependence of $d_{33}$ on the applied electric field due to intrinsic domain reversal (i.e. no contributions such as ferroelastic motion or field-induced phase transitions) along the [001] direction of the film, using [14]:

$$d_{33}(E) = 2\overline{Q}P(E)\varepsilon_{33}(E), \tag{1}$$

where $P$ (= $P_3$) is the polarization and $\varepsilon_{33}$ is the relative dielectric constant. $\overline{Q}$ is an effective electrostrictive coefficient that accounts for the clamping effect of the substrate. This phenomenological relation states that the field dependence of the *$d_{33}$* coefficient is principally governed by the field dependence of the polarization as well as the dielectric susceptibility. The main features of the AFE *$d_{33}$* loop are consistent with the *P-E* loop (Fig. 3) and the *$\varepsilon_{33}$-E* curve (Fig. 4 inset). The black arrow marks the region where electric-field induced transition from the antiferroelectric to ferroelectric state is taking place. The eventual maximum in the *$d_{33}$* loop is a consequence of the increase in the switching polarization in the ferroelectric state with a sharp positive change of the slope of the *P-E* loop and increase in the dielectric susceptibility at that point. As the applied electric field increases further, an inflection point is reached beyond which the net increment in switchable polarization for given increase in applied field begins to decrease. As a consequence, the dielectric susceptibility begins to decrease, and hence also brings down the *$d_{33}$* value (the red arrow). At higher electric field, the polarization is fully switched and now similar to a "fully saturated state" in a standard ferroelectric. Here, the drop in the dielectric susceptibility with increasing electric field dominates the shape of the *$d_{33}$* loop, which shows a downward slope (the green arrow).

The present results show that $Sm^{+3}$ substituted BFO has unique properties compared to more popular lanthanum (La)[16] or neodymium (Nd)[3,4] doped BFO. It was reported that 5% substitution of La (or Nd) in BFO thin films resulted in reduction of the crystal anisotropy by more than 10% accompanied by a drastic reduction in the switchable polarization (from 79 µC/cm$^2$ to 40 µC/cm$^2$).[3] However, no distinct increase in the electromechanical properties or the occurrence a complex domain structure was reported. Suchomel and Davies have shown that any advantage gained by introducing a structural phase transition via doping with a non-lone pair cation is offset by the dramatic reduction in polarization due to the nonpolarizable dopant, as is the case for $La^{+3}$.[9] Further, $Sm^{+3}$ (1.24 Å) has a much smaller ionic size than $La^{+3}$ (1.38 Å). Thus, the Goldschmidt tolerance factor[10] for the $Sm^{+3}$ doped BFO is less than 1, and lower symmetry structures such as triclinic are more likely at the MPB. The presence of a low symmetry phase at the MPB is expected to enhance the piezoelectric properties as the polarization vector is no longer constrained to lie along a symmetry axis but instead can rotate within a suitable plane.[7, 17]

Ravindran et al [18] have recently shown through density functional calculations that BFO undergoes a rhombohedral (R3c) to orthorhombic (Pnma) structural transition under pressure. Because of the small ionic radius, chemical pressure due to the continuous $Sm^{+3}$ doping can perhaps be viewed as resulting in the similar lattice instabilities. The appearance of the triclinic phase at the boundary is also consistent with the fact that the dopant substituted structure is lower in symmetry than the rhombohedral BFO or the orthorhombic $SmFeO_3$ and is a subgroup of either structure.

We are grateful to K. M. Rabe for valuable discussions. H. Oguchi is acknowledged for XRD measurements. This work was supported by access to the Shared Experimental Facilities of the UMD-NSF-MRSEC (DMR 0520471), NSF

DMR 0603644, and ARO W911NF-07-1-0410. The work was also supported by the W. M. Keck Foundation. The work at UNSW is supported partially by an Australian Academy of Science Travel Fellowship and DEST International Science Linkage Scheme. The work at ANL was supported by the U. S. Department of Energy, Office of Science, Office of Basic Energy Sciences, under Contract No. DE-AC02-06CH11357.

**Figure Captions**

Figure 1.

X-ray diffraction and structural phase diagram of $Bi_{1-x}Sm_xFeO_3$ (a) $\theta$- $2\theta$ scan from 43º to 48° across a composition spread (6 mm long) on [001] $SrTiO_3$ shows continuous change in the film lattice constant. The red line traces the (002) peak of $Bi_{1-x}Sm_xFeO_3$ except for the dotted region where the peak is obscured by the substrate peak. (b) Two-dimensional diffraction images taken from x = 0 to x = 0.3 reveals appearance of the (1/2 0 2) and (-1/2 0 2) spots indicating occurrence of cell doubling at x ≈ 0.14. The structural evolution in $Bi_{1-x}Sm_xFeO_3$ is shown below going from rhombohedral $BiFeO_3$ (unit-cell in red) to orthorhombic $SmFeO_3$ (unit-cell in blue). The triclinic structure observed at the morphotropic phase boundary (MPB) x ≈ 0.14 can be viewed as having a doubled distorted rhombohedral structure or a distorted (pseudo) orthorhombic structure. Beyond x ≈ 0.27, the compound was found to be paraelectric (PE).

Figure 2.

High resolution plan-view transmission electron microscopy image (left) of the composition at the morphotropic phase boundary (MPB, x = 0.14). The triclinic nanodomains 20-50 nm in size are observed with twin boundaries (yellow dotted lines). The arrows denote the (010) direction of the domains. The selected area electron diffraction of this region (right top) is consistent with superposition of diffractions from nanodomains of the triclinic structure in 3 different orientations I, II, and III (right bottom). The orientations here are indexed for the triclinic structure.

Figure 3.

Continuous change in the ferroelectric hysteresis loops (5 kHz) is observed as a function of changing composition. Three representative compositions were selected for display for clarity. The relatively high polarization of ≈ 70 μC/cm$^2$ is maintained as we go from BiFeO$_3$ to Bi$_{0.86}$Fe$_{0.14}$O$_3$. Beyond MPB and before x ≈ 0.27, the material displays antiferroelectric (AFE) characteristics (as seen for x = 0.16).

Figure 4.

(a) Dielectric constant ($\varepsilon_{33}$) and tan δ measured at 1 MHz (zero-bias). $\varepsilon_{33}$ shows a broad peaking behavior with a maximum at x = 0.14, while the loss tangent remains relatively low. The curve is a guide to the eye. The inset shows the $\varepsilon_{33}$ vs. electric field for the AFE composition at Bi$_{0.84}$Sm$_{0.16}$FeO$_3$.

Figure 5.

Piezoelectric properties of Bi$_{1-x}$Sm$_x$FeO$_3$ (a) $d_{33}$ loops for Bi$_{0.87}$Sm$_{0.13}$FeO$_3$ and Bi$_{0.86}$Sm$_{0.14}$FeO$_3$. The high value of remanent $d_{33}$ for Bi$_{0.86}$Sm$_{0.14}$FeO$_3$ is comparable to that of PZT thin films of the same thickness at MPB. (b) High-field piezoelectric coefficient ($d_{33}$) determined from the piezoelectric hysteresis loops measured as a function of composition. The average value obtained for the positive and negative field is plotted. The $d_{33}$ constant shows a sharp peak at x = 0.14. The curve is a guide to the eye. The inset shows the Rayleigh analysis measurements for the MPB composition sample. (c) Antiferroelectric $d_{33}$ loop is observed for Bi$_{0.84}$Sm$_{0.16}$FeO$_3$.

**Figure 1**

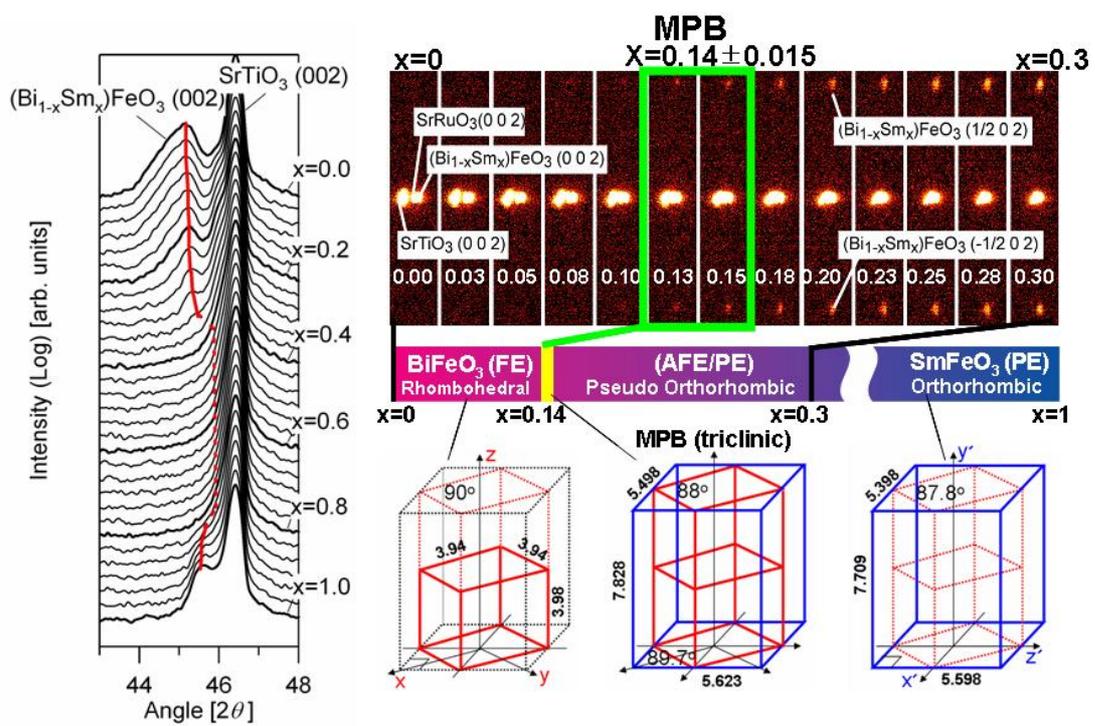

**Figure 2**

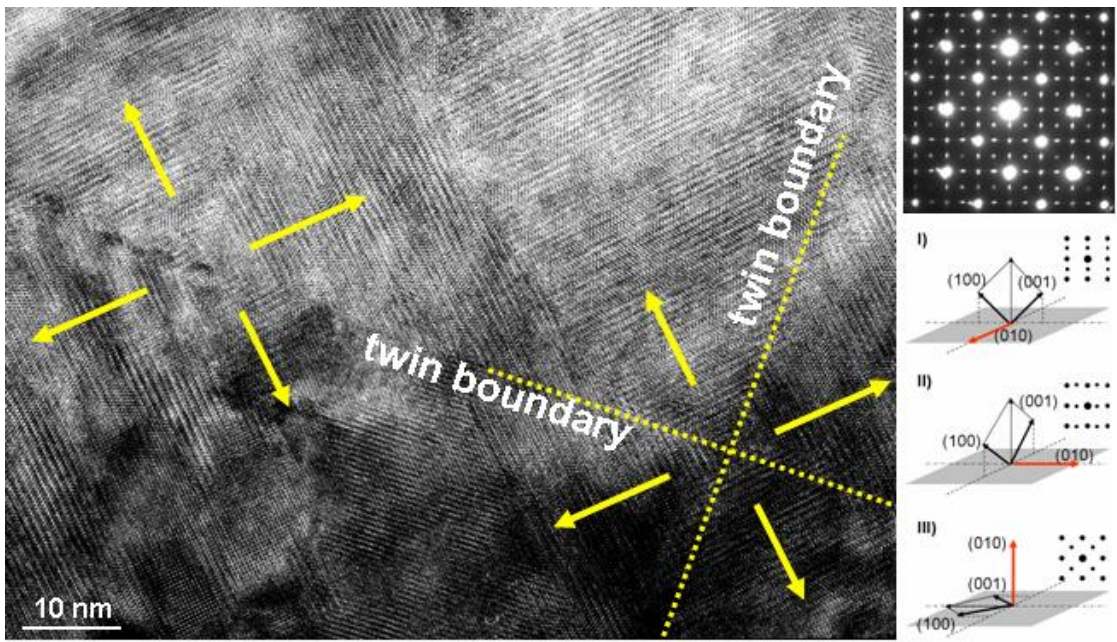

**Figure 3**

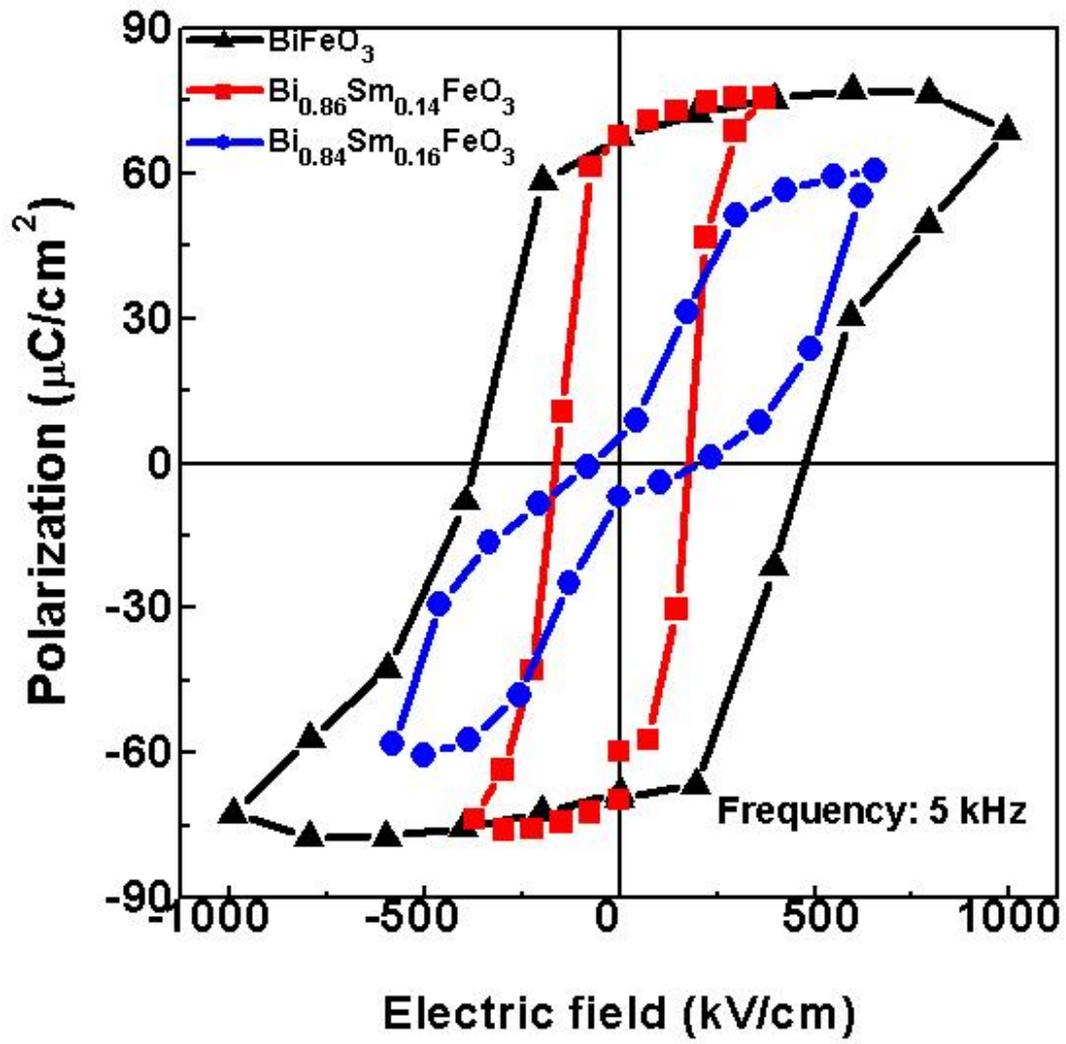

**Figure 4**

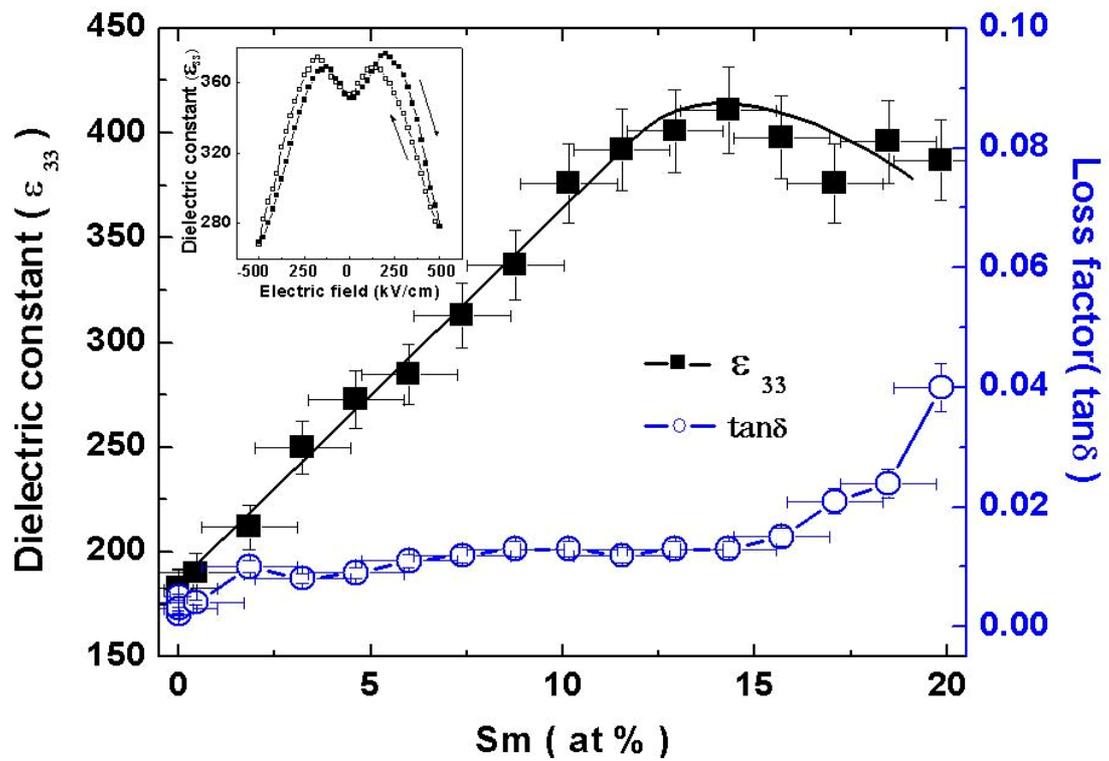

**Figure 5**

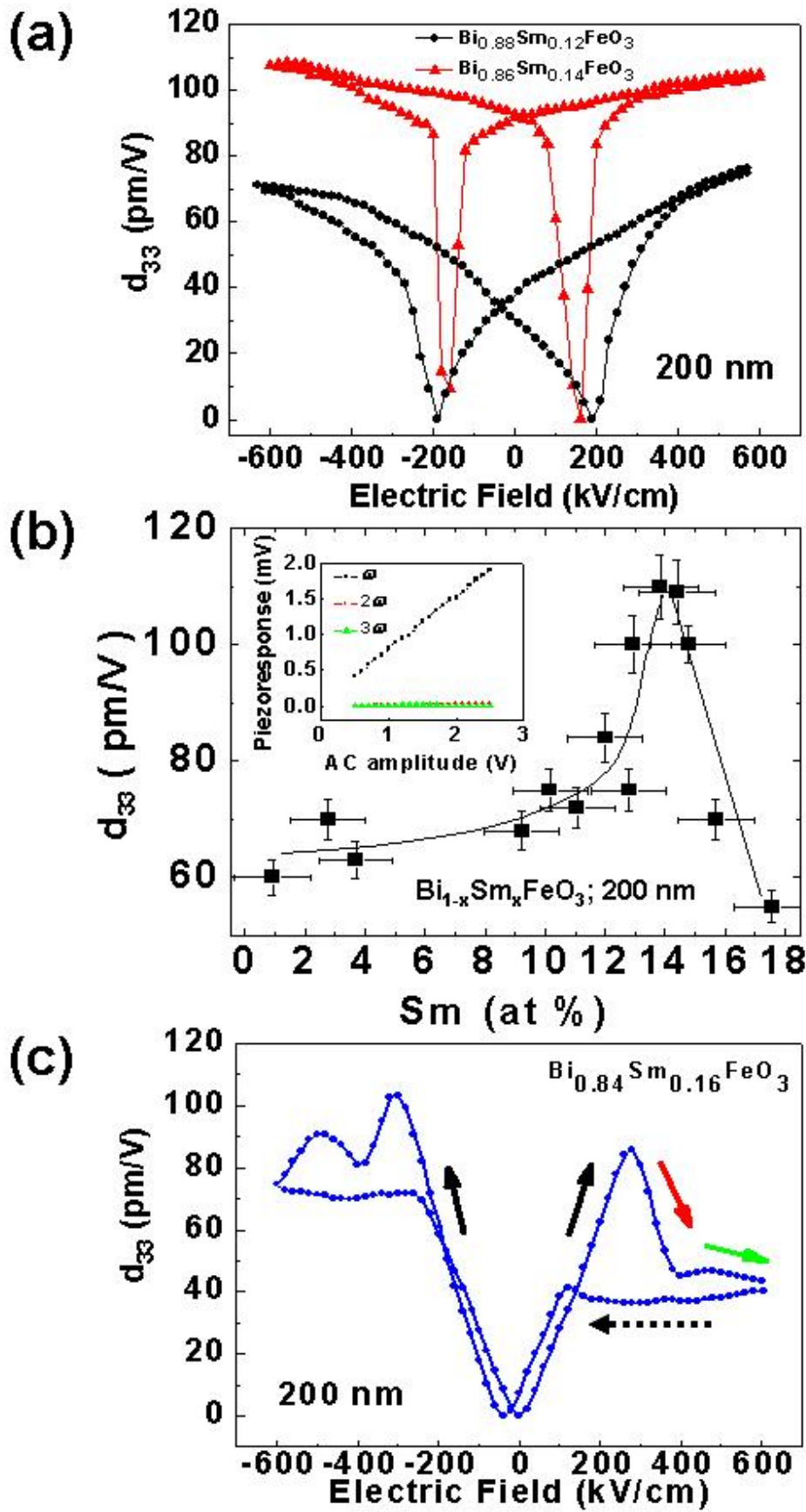